\documentstyle[12pt,prl,aps]{revtex}

\draft
\begin{document}

\title{
Superconductivity in the Two-Band Hubbard Model in Infinite D:
an Exact Diagonalization Study}

\author{Werner Krauth$^*$ and Michel Caffarel$^{**}$}
\address{
$^{*}$ CNRS-Laboratoire de Physique Statistique de l'ENS\\
24, rue Lhomond; F-75231 Paris Cedex 05; France\\
e-mail: krauth@physique.ens.fr\\
$^{**}$ CNRS-Laboratoire de Physique Quantique$^1$\\
IRSAMC, Universit\'{e} Paul Sabatier\\
118, route de Narbonne; F-31062 Toulouse Cedex; France\\
e-mail: mc@tolosa.ups-tlse.fr\\}
\date{June 1993}
\maketitle
\begin{abstract}
We apply an exact diagonalization method to the
the infinite-D two-band Hubbard model. The method is
essentially exact for the calculation of thermodynamic properties for all
but the smallest frequencies
and yields a resolution unavailable in Monte Carlo calculations.
We establish the instability of the normal state
with respect to singlet superconductivity at small frequencies at
small doping, the regime of relevance for High-$T_c$
superconductors.
We also present evidence for the existence of an instability towards
 triplet superconductivity in
the large doping regime $n \sim 2$.
\end{abstract}
\newpage
The discovery of high-temperature superconductivity has enormously
increased the interest in strongly correlated electron systems.
Superconductivity in its various guises has been searched for in
practically all available models. Of particular interest
are models with several electronic bands, such as the three-band model
of Emery \cite{EM} and of
Varma, Schmitt-Rink and Abrahams \cite{VSA}. These have been proposed
as minimal models for the $CuO$ planes in the new materials.
In 2D, the three-band model has been been actively studied
numerically, but even the most ambitious Quantum Monte Carlo calculations
\cite{DMH} have up to now given only inconclusive evidence.

Given the enormous complexity of the full  problem in two or three
dimensions it is natural  to bring to bear
the methods of the - now well established -  infinite dimensional
approach on this problem \cite{VOL}.
As for classical statistical mechanics systems, the limit of infinite
dimensions constitutes a self-consistent mean field theory
in which the spatial fluctuations are frozen and where
the infinite-D many-body system is reduced to a self consistent
single site problem.
It is of considerable interest to clarify whether  such
models with several bands are able to show superconductivity
even (or: at least) in
large D, {\it i.e.} in the absence of spatial ({\it e.g.} antiferromagnetic)
fluctuations. This is the aim of the present paper.

We investigate in detail the infinite-D two band Hubbard model,
which was proposed in a recent paper \cite{GKK}
and  studied with an extension of the
Hirsch-Fye Quantum Monte Carlo (QMC) algorithm \cite{HF}.
This method proved to be
adequate for the investigation of the normal state, and several very
interesting metal-to-insulator transitions were identified.
However, the QMC algorithm
did not allow a controlled investigation of the superconducting state
or of the instability of the normal solution. The main result was that,
at the temperatures accessible to the QMC method,
the superconducting susceptibilities
were large in two different regimes, the small doping regime (of
relevance for high-$T_c$ superconductors) and a regime with density
$n$ close to $2$ for intermediate values of the Coulomb interaction.

In this paper we take up these two regimes ($n \stackrel{>}{\sim} 1$
and $n \sim 2$). We are interested in the normal state exclusively as
a starting point for a linear stability analysis and investigate
the regime close to the normal solution.
Before presenting our results we will give a brief explanation of
our exact diagonalization
method ({\it cf} \cite{CK}) as applied to the problem of superconductivity.
Just as for the Hubbard model \cite{CK}, this method is
far superior to the Quantum Monte Carlo method and
allows us to decrease by at least an order of magnitude
the limiting temperature (smallest accessible frequency).

The $(CuO)_d$-type which is the object of our study
is defined by the Hamiltonian \cite{GKK}
\begin{equation}
{\cal H}\,=\,
- \sum_{i\in D,j\in P,\sigma} t_{ij} d^{+}_{i\sigma} p_{j\sigma} + h. c.
+ \epsilon_p \sum_{j\in P,\sigma} p^{+}_{j\sigma} p_{j\sigma}
+ \epsilon_d \sum_{i\in D,\sigma} d^{+}_{i\sigma} d_{i\sigma}
+ U_d \sum_{i\in D} n^d_{i\uparrow} n^d_{i\downarrow}
\label{H}
\end{equation}
where the hopping is scaled as $t_{ij} \sim 1/\sqrt{2z}$ (z is the
connectivity of the lattice). In eq.
(\ref{H})
$(d_{\sigma},p_{\sigma})$ represent two atomic orbitals
on different sublattices $(D,P)$ of a
bipartite lattice with $z \rightarrow \infty$ which,
for the purposes of the present paper, is taken
to be the infinitely connected Bethe Lattice.
The repulsive Coulomb interaction acts
only on the  ('copper') orbital $d_{\sigma}$,
while the ('oxygen') orbital $p_{\sigma}$ is uncorrelated.
In the standard Nambu notation,
$\Psi_d^{+} \equiv (d_{\uparrow}^{+}, d_{\downarrow})$ (equivalently for
$\Psi_p$) the d-orbital Green's function can be written
as a $2 \times 2$ matrix
\begin{equation}
{\bf D}(\omega) \equiv  -T<\Psi_d(\omega) \Psi_d^{+}(\omega)>\,=
\left(
\begin{array}{ll}
G_d(\omega)\,&F_d(\omega)\,\\
F_d(\omega)^*\,&-G_d(- \omega)\,\\
\end{array}
\right)
\label{gfsup}
\end{equation}
and the self-consistency equations for the Green's functions are given
by \cite{GKK}
\begin{equation}
\begin{array}{ll}
{\bf D_0}^{-1}(i\omega_n) = i\omega_n + (\mu-\epsilon_d)\sigma_3
-t_{pd}^2 \,\sigma_3 {\bf P} (i\omega_n) \sigma_3\,\\
{\bf P}^{-1}(i\omega_n) = i\omega_n + (\mu-\epsilon_p)\sigma_3
-t_{pd}^2 \,\sigma_3 {\bf D} (i\omega_n) \sigma_3
\end{array}
\label{scs}
\end{equation}
(note that $D_0$ and $D$ are $2 \times 2$ matrices and that ${\bf D_0}^{-1}$
denotes the matrix inverse).

The Green's function $D(i\omega_n)$ and the
bath Green's function $D_0(i\omega_n)$ are connected  by the single-site
action
\begin{equation}
{\cal S}_{sup}\,=\,U_d\,\int^{\beta}_{0} d\tau\, n_{d\uparrow}(\tau)
n_{d\downarrow}(\tau)
- \int^{\beta}_{0} d\tau \int^{\beta}_{0} d\tau'
\Psi_d^{+}(\tau) {\bf D_0}^{-1}(\tau-\tau') \Psi_d(\tau')
\label{action}
\end{equation}
This eq. (\ref{action}) is at the base of the Quantum
Monte Carlo approach to the
solution of the $d=\infty$ equations, which consists in a
decoupling of the interaction term in $U_d$ and a proper
discretization of the integrals.
Alternatively, the Green's functions $D(i\omega_n)$ and $D_0(i\omega_n)$
may be viewed  as impurity Green's functions of an
effective Anderson model \cite{GK} which, in the presence of a
superconducting medium, is given by
\begin{equation}
H_{AM} = \sum_{\sigma,l=1}^{n_s} \epsilon_{l} a^{+}_{l\sigma}a_{l\sigma} +
U n_{1\uparrow}n_{1\downarrow} + \sum_{\sigma,l=2}^{n_s}
[V_{l}a^{+}_{l\sigma}a_{1\sigma} + h.c] + \sum_{l,k=1}^{n_s}
\phi_{lk} [a^{+}_{l\uparrow} a^{+}_{k\downarrow} + h.c.]
\label{impurity}
\end{equation}
In eq. (\ref{impurity}) we have used a compact notation with
a site index $l$, $1 \leq l \leq n_s$, in which $l=1$ denotes the impurity.
The matrix $\phi$ provides explicit pairing terms between all the sites.
The matrix is to be taken antisymmetric for the calculation of
the triplet sector and symmetric in the
singlet sector.
Of course
a strictly self-consistent solution based on the correspondence
between eq. (\ref{action}) and
eq. (\ref{impurity}) will only
be possible in the limit of an infinite
Anderson model $n_s = \infty$.

The exact diagonalization method consists in approximating the
function $D_0(i\omega)$ in eq. (\ref{scs}) by the
Green's function of
a (superconducting) impurity model eq. (\ref{impurity})
with a {\it finite}, and even small number  $n_s$ of sites.
Explicitly, $D_0^{And}(i\omega)^{-1} = ( i \omega -H)$
with $H$ the single particle Hamiltonian (U=0) corresponding to eq.
(\ref{impurity}).
[In the normal state, in which $\phi = 0$, the Green's function is
given by the standard formula
$G_{0\;d} (i\omega)^{-1} = i \omega -\epsilon_d - U_d/2 -
\sum_{k=2}^{n_s}V_k^2/(i \omega -\epsilon_k)$].

In practice,
we use  a conjugate gradient method to determine the parameters
$V_k, \epsilon_k$, and $\phi_{ij}$ minimizing the function
\begin{equation}
           E= \sum_{n=0}^{n_{max}} | D_0(i \omega_n)-D_0^{And}
(i \omega_n) |/(n_{max}+1)
\label{fitfun}
\end{equation}
Here the $\omega_n$ denote the smallest Matsubara frequencies.
Even at zero temperature we use $\omega_n= (2 n + 1)\pi/\beta$ with
$\beta $
a {\it fictitious} temperature. This provides a lower frequency
cutoff. All the results we present do not critically depend on the
specific choice of the fitting function, and on $n_{max}$.

For $n_s \leq 6$ we are able to diagonalize exactly the
Hamiltonian eq. (\ref{impurity}), with a full calculation of the spectrum
and of the eigenvectors. At temperatures accessible to QMC, the
Green's functions,
susceptibilities, {\it etc}  can then be
compared with the extrapolated results of the
Monte Carlo algorithm to very high precision.
Since we are mainly interested in Green's functions in the zero-temperature
limit, we may use the Lancz\`os method \cite{LAN} and are then
able to increase noticeably the size of the cluster. For the
full superconducting Hamiltonian eq. (\ref{impurity}) values up
to  $n_s \sim 8$ may
be handled without problems on a work station \cite{foot1}.

The algorithm which we have just sketched is then iterated to convergence.
Once this is reached we are able to estimate the quality of
the calculation, which is entirely determined by the
agreement of the two Green's functions $D_0$ and $D_0^{And}$.
(Note that this is
an {\it intrinsic} criterion).
The  agreement between the two functions
has been found to be amazingly good. We insist that
the fit of $D_0$ by $D_0^{And}$ is the only non-exact part of
our procedure and seems to be a much less violent approximation
than those introduced by the Trotter breakup
(discretization of the integrals
in eq. (\ref{action})) and by the stochastic noise of the Monte Carlo
procedure.

An example of the excellent quality of the solution
is given by the normal state results \cite{normal}
at zero temperature presented
in fig. 1. Here,
the functions $G_{d0}(i\omega)$ and $G_{d0}^{And}(i\omega)$ are displayed.
The 'effective' temperature is $\beta=250$, $U_d=8$, and the density
corresponds to
the lightly doped regime of the two-band model. At $n_s=8$, the
differences are below the resolution of the graph with a maximum difference
on the order of $10^{-2}$.
At fixed value of $n_{max}$ (here taken to be 64),
we have noticed a {\it systematic} improvement of the quality of fit,
expressed by a value of $E$ in eq. (\ref{fitfun}) which decreases
by a factor of $\sim 4$ each time one more site is added \cite{footqual}.
As an illustration
of this improvement we show in the insets of fig. 1 the
self-consistent solutions at $n_s$=4,5,6,7, and 8.
Above $i \omega \sim .5$ the Green's functions
are strictly identical.

We now consider the stability analysis of the normal state solution.
A possible way of studying this stability is to
calculate the pairing susceptibility. An alternative way
used here is to establish
the stability properties of the solution by introducing small terms
$\phi_{ij}$ in the Hamiltonian eq. (\ref{impurity}), and following
the evolution under subsequent iterations
\cite{GKK}. Under such conditions, the normal
state solutions in fig. 1 very quickly acquire non-zero values of
$F(\omega)$, which indicate a superconducting instability.
More rigorously, and in order to study quantitatively the effects of
increasing $n_s$, we may calculate the largest eigenvalue, and the
corresponding eigenvector
of the matrix
$\partial F(i\omega)^{n+1}/
\partial F(i\omega)^{n}$
close to the normal state, where the
superscripts on the $F's$ indicate two subsequent iterations of the
self-consistency loop. This involves a simple rescaling of the
$\phi$ at each iteration. We have done such calculations, which correspond
to the well-known procedure of extracting the largest eigenvalues
and eigenvectors of a matrix with the 'power method', starting
from the parameters in fig. 1, for $5 \leq n_s \leq 8$. We are
able to identify a linear regime at small $\phi$, with
the largest eigenvalue always of the order $\lambda_{max} \sim 2$.
The corresponding
(rescaled) eigenvectors for $n_s = 6,7,8$ are plotted in fig. 2.
Clearly, the agreement between these completely independent curves is
excellent. We have checked this result in a variety of ways (by changing
the effective temperature, the precise form of the function
used in eq. (\ref{fitfun}), and the doping).
This leads us to the conviction that the normal state
solution of the $d=\infty$ model at small doping is indeed unstable
with respect to singlet superconductivity.

We have also studied the point investigated previously \cite{GKK}, {\it i.e.}
values of the physical parameters corresponding to a total density
of $n \sim 2$, where the Hubbard interaction is just large enough to
create a large overlap
between the upper Hubbard band of the
$d$-level and the $p$-level band. There our evidence for singlet
superconductivity is very limited
(at least for frequencies larger than $\sim 1/200$). However, we have
on that point found very clear evidence for superconductivity in the
triplet sector.
Following the procedure outlined above, we find consistently at small
'effective' temperature that any small
terms in $\phi$, in addition to the normal state solution, blow up at
a rate which corresponds to a largest eigenvalue of $\sim 1.8$  of the matrix
$\partial F(i\omega)^{n+1}/
\partial F(i\omega)^{n}$. In fig. 3 we show the zero-temperature
normal state solution for $U_d=4.5$, $\mu=\epsilon_p - \epsilon_d=4$,
calculated
on a grid of points corresponding to an effective temperature of
$\beta=200$. The inset shows the most unstable eigenvector in the
triplet sector. Its corresponding eigenvalue is $\lambda_{max} = 1.75$
\cite{footprec}. Superconducting order of this kind has been first
proposed by
Berezinskii \cite{BER} in the context of $^3$He, and, very recently by Coleman,
Miranda and Tsvelik \cite{CMT} for heavy-fermion superconductors.

In conclusion, we have studied the infinite dimensional two-band Hubbard model
with an exact diagonalization method, which has given very strong evidence for
a superconducting instability at low temperature.
Given the excellent fit of the Green's functions, and the smallness
of the finite size effect (dependence on $n_s$), it seems to us to be
difficult
to escape the conclusion  that the two-band model is indeed
superconducting, with a strongly frequency-dependent order parameter.
We have taken every effort to check the programs  (exact
diagonalization and Lancz\`os) against each other and against
the extrapolated Quantum Monte Carlo results at sufficiently high temperature.
Close to the normal state solution we have also not been plagued
by possible multiple solutions of the minimization in eq. (\ref{fitfun}).
In fact, we have also been able to find perfectly converged
solutions of the fully superconducting phase \cite{KC2}, but
there the extrapolation with $n_s$, and the possible
problem of multiple solutions are more critical. Therefore, the
linear stability analysis presented here is the most convincing
argument in favor of a superconducting instability of the
two-band Hubbard model which we are able to give at the present time.

Finally, as has been discussed elsewhere
\cite{CK}, the method does for the time being not allow
a precise calculation of the densities of states, and a renormalization
group procedure is clearly called for. Such a method would allow
the calculation of regular and superconducting densities of state.
It would also allow calculations at finite temperature with much larger
precision. The search for a numerical  renormalization
group procedure for $D=\infty$ is
in our opinion an extremely exciting challenge to take up.

\acknowledgments
We acknowledge helpful discussions with
J. Bellissard, A. Georges, G. Kotliar, D. Poilblanc and T. Ziman.
This work was supported by DRET contract $n^o\;921479$.

\newpage
{\bf Figure Captions}
\begin{enumerate}
\item{}
\label{fig1}
Zero temperature imaginary-time functions $G_{d\;0}(i\omega)$ and
$G_{d\;0}^{And}(i\omega)$
(real parts: upper) (imaginary part: lower) {\it vs} $\omega$
$U_d=8$,
$\mu=\epsilon_p-\epsilon_d=4$ as calculated
with the Lancz\`os algorithm with $n_s=8$ sites. Inset:
Real part of the functions for $n_s=4,5,6,7$ and $=8$.
If $G_{d\;0}^{and}(i\omega)$ and
$G_{d\;0}(i\omega)$ were exactly equal this would be the exact solution.
Note the excellent fit of the two quantities for $n_s=8$, and the
systematic improvement with increasing number of sites. A quasi-exact
solution is obtained at $n_s=$8.

\item{}
\label{fig2}
Largest eigenvector of the matrix $\partial F(i\omega)^{n+1}/
\partial F(i\omega)^{n}$ close to the normal state solution of fig. 1.
for $n_s=6,7$ and $8$ (singlet sector).
The corresponding eigenvalues are $\lambda_{max} \sim 2$ in all three
cases.
\item{}
\label{fig3}
Zero-temperature Green's function $G(i \omega)$ in the normal state
at $U_d=4.5$, $\mu=\epsilon_p -\epsilon_d=4$
at $n_s=7$ (effective $\beta=200$, $n_{max}=64$). The misfit between
$G_{d\;0}(i\omega)$ and $G_{d\;0}^{and}(i\omega)$ is
given by $E= 1.5\times 10^{-4}$ (!) ({\it cf.} eq. \ref{fitfun}) and
the maximum difference between the two functions is
$7 \times 10^{-4}$.
The inset shows the most unstable eigenvector in the triplet sector
at $n_s=7$. The corresponding eigenvalue is $\lambda_{max}=1.75$.
\end{enumerate}

\begin{references}
\bibitem{foot1}
In practice, the Hamiltonian eq. (\ref{impurity}) is rewritten in
terms of Nambu spinors. The pairing terms are then transformed into
spin-flip operators, and conserve the number of up-particles + down-holes.
This symmetry has been implemented.
\bibitem{normal} This means that the anomalous Green's function
is forced to be $0$ and that a self-consistent solution for
$G_d(i\omega_n)$ is searched for. $F_d=0$ is always a possible
solution of the self-consistency equation eq. (\ref{scs}).
\bibitem{footqual}
This observation (which would indicate exponential convergence in $n_s$)
is our main argument in favor of the algorithm. It seems rather unlikely that
we would discover new physics at values  of $n_s$ much larger than the
ones treated in the presence of both an excellent fit and of extremely
rapid convergence.
\bibitem{footprec} In the triplet case we have very good convergence
of the eigenvalues as a function of $n_s$, but the eigenvalues -although
qualitatively the same -
show some quantitative differences at small frequencies. Larger
simulations are probably called for in order to settle this point.
\end{references}
\end{document}